\documentclass{iosart2c}
\usepackage[T1]{fontenc}
\usepackage{times}%
\usepackage[utf8]{inputenc}
\usepackage{epsfig}
\usepackage{amsmath, amssymb, amsthm}
\theoremstyle{definition}
\newtheorem{theorem}{Theorem}[section]
\newtheorem{definition}[theorem]{Definition}
\usepackage{dcolumn}
\usepackage{endnotes}
\usepackage{graphics, graphicx}
\newcolumntype{d}[1]{D{.}{.}{#1}}
\begin{document}
\begin{frontmatter}  
%\pubyear{0000}
%\volume{0}
%\firstpage{1}
%\lastpage{1}
%\pretitle{}
\title{Design of generalized fuzzy multiple deferred state (GFMDS) sampling plan for attributes}
\runningtitle{Design of GFMDS sampling plan for attributes}
\author[A]{\fnms{Julia} \snm{Thampy Thomas}} and
\author[B]{\fnms{Mahesh} \snm{Kumar} \thanks{Corresponding author. E-mail: mahesh@nitc.ac.in}}
%write corresponding as footnote
\runningauthor{J. T. Thomas and M. Kumar}
\address[A]{Department of Mathematics, National Institute of Technology, Calicut, India 673601\\
E-mail: julia.maths1.61@gmail.com}
\address[B]{Department of Mathematics, National Institute of Technology, Calicut, India 673601\\
E-mail: mahesh@nitc.ac.in}

\begin{abstract}
A sampling plan is a pilot tool for a supply and demand chain quality check strategy. These plans proved to be economically viable for the quality inspection processes but the uncertainty in the plan parameters challenged the reliability of the application of traditional acceptance sampling plans. This study proposes a generalized fuzzy multiple deferred state (GFMDS) sampling plan for attributes that consider the ambiguity in determining the exact value of the percentage of defectives in a batch. The performance measures have been derived and the plan is designed in terms of a minimum average sample number. A comparison study is done over the existing fuzzy acceptance sampling plans for attributes and a pertinent observation is made regarding the efficiency of the GFMDS scheme. The effect of inspection errors on the sampling procedure is analyzed and the drop in the acceptance criteria of the plan is observed corresponding to the intensified inspection errors. Several numerical examples are presented to validate the results. 
\end{abstract}

\end{frontmatter}

\section{Introduction}
Logistics and supply chain management are integral parts of manufacturing systems and the quality assurance of the products is often challenging. Acceptance sampling plans are carried out extensively in industrial quality assurance processes. The lower cost of inspection and minimal risks makes acceptance sampling plans more attractive. The quantitative measurements of the inspection process are handled by acceptance sampling plan for variables whereas the qualitative nature is inspected by attribute sampling plans. Lowering inspection costs depends on the proportion of defectives and the sample size. Facilitating an acceptance sampling plan with the lowest average samples inspected is an efficient way to ensure quality control from an economic perspective.

The ease of application of attribute sampling plans in industries made them more popular than variable sampling plans for which more information is needed. The history of acceptance sampling plans for attributes dated long back when a method of sampling inspection was introduced \cite{dodge1929method}. Single sampling plans (SSP) and double sampling plans (DSP) were the pioneers in acceptance sampling schemes (see, \cite{dodge1941single}) and eventually the limitations of these sampling schemes were identified and rectification is carried out. The information furnished by the current sample was the determining factor in the sampling plans designed initially. The chain sampling scheme used cumulative information of several samples to accept or reject a lot \cite{dodge1977chain}. The complexity of chain sampling is relaxed when a new acceptance sampling scheme was designed where the stages of inspection were inter-dependent (see \cite{wortham1970dependent}), which later evolved through the design of multiple deferred state (MDS) sampling plans as seen in \cite{wortham1976multiple}. The stages of attribute MDS acceptance plan are acceptance, rejection, or conditional acceptance of the lot. The advantage of MDS plans over SSP and DSP is the reduced sample size. In the MDS plan, the sample information from the previous samples is employed to conclude the decision on the lot submitted. The selection of optimal parameters for MDS plans is studied in \cite{soundararajan1990construction} and \cite{kuralmani1992selection}. The determination of plan parameters concerning the acceptable quality level (AQL) and the limiting quality level (LQL) is studied in \cite{govindaraju1993selection}. Later, the MDS plan is used to develop an attribute control chart (see \cite{aslam2015new}). Variable sampling schemes using the MDS method were also developed and gained relevance over the existing plans (see \cite{balamurali2007multiple}, \cite{aslam2014multiple}, \cite{wu2017developing}, \cite{yan2016designing} and \cite{aslam2021determination}).

The crisp nature of the fraction of non-conformity's ($p$) is not assured in real-world supply chain models. This ambiguity often arises when a linguistic value is used to mention the number of defectives. The development of acceptance sampling plans on the ground of fuzzy logic and applications proved more applicable to these cases. Fuzzy set theory is used for the mathematical modeling of epidemiology, biostatistics, reliability engineering, statistical quality control, and so on. Fuzzy optimization is used to develop a single sampling plan \cite{chakraborty1992class}. Risks and quality attributes were considered as fuzzy quantities to design an acceptance sampling plan by attributes \cite{grzegorzewski2001acceptance}. A fuzzy SSP is developed by \cite{sadeghpour2011acceptance} where the plan parameters were considered as fuzzy numbers. Further, developments in fuzzy acceptance sampling plans for attributes can be seen in \cite{turanouglu2012fuzzy}, \cite{kahraman2016fuzzy} and \cite{isik2021design}. Consequently, the uncertainty in the fraction of defectives in the MDS sampling plan for attributes is addressed in \cite{afshari2017construction} and \cite{afshari2017designing}. A variable MDS plan with fuzzy parameters is developed in \cite{afshari2018fuzzy}.

The chances of inspection errors are often ignored in the quality control processes. However, no sampling schemes proved perfect when applied to a supply chain where the assumptions of sampling schemes undergo fluctuations from their theoretical framework. The effect of classification errors in acceptance sampling schemes was studied in \cite{collins1973effects} where the SSP was modified in this respect. Later, many plans were rephrased by considering the chances of misclassifications of the submitted lot for attribute sampling plans (see, \cite{fard1993analysis}, \cite{ferrell2002design}, \cite{govindaraju2007inspection}, and \cite{anderson2001acceptance}). The effect of measurement errors on fuzzy acceptance sampling plans were also studied \cite{baloui2011inspection}. The case with the MDS plan was analyzed in \cite{afshari2018effects} and its fuzzy counterpart was studied in \cite{afshari2017fuzzy}. The results proved a notable drop in the probability of acceptance when the errors are identified. 

When the limitations of MDS sampling plans were identified, the assumptions of the MDS sampling scheme were generalized to consider measurement data \cite{bhattacharya2020generalized}. The efficiency of this plan over the existing MDS plan versions was discussed. The economic advantage of the plan was considered in \cite{aslam2021determination} which significantly contributed to the reduction in total cost for supply chain models. The reduction in average sample number is the key to obtaining optimal parameters for generalized MDS plans which eventually contributed to low inspection costs. This research is highly motivated by the extensive applications of fuzzy logic to the probability theory and quality assurance models. The objective of this article is to develop a generalized fuzzy multiple deferred state acceptance sampling plan where the limitations of the crisp nature of $p$ in a generalized MDS plan can be resolved. The advantage of the proposed plan is discussed by comparing the average sample number of the existing fuzzy acceptance sampling plans. The organisation of this paper is as follows. The forthcoming section carries some preliminaries of fuzzy sets. The assumptions and a systematic design of the GFMDS plan is given in Section 3 where the measures of performance are derived. Section 4 deals with the determination of suitable parameters for the GFMDS plan and the plan is extended to generalised fuzzy numbers as seen in Section 5. Section 6 discusses the efficiency of the sampling plan under consideration. Further, the GFMDS plan is modified with consideration of inspection errors and presented in Section 7. Conclusions is mentioned in the Section 8. 

\section{Some Preliminaries} 
This section presents some fundamentals related to fuzzy numbers.
\begin{definition}\cite{dubois1978operations}
The fuzzy subset $\tilde{X}$ of real line $\mathbb{R}$ with membership function $\mu_{\tilde{X}}:\mathbb{R}\rightarrow[0,1]$ is a fuzzy number if and only if
\begin{enumerate}
    \item $\exists x_0 \in \mathbb{R} s.t. \mu_{\tilde{X}}(x_0)=1$
    \item $\forall x_1, x_2 \in \mathbb{R}$ and $\forall \lambda \in [0,1]$ implies that $\mu_{\tilde{X}}[\alpha x_1 +(1-\alpha)x_2] \geq \min \{\mu_{\tilde{X}}(x_1),\mu_{\tilde{X}}(x_2)\}$
    \item for each given $x \in \mathbb{R}$ and $\epsilon>0, \exists \delta >0$ s.t. $\forall z \in \mathbb{R}, |x-z|<\delta$ implies that $\mu_{\tilde{X}}(z)<\mu_{\tilde{X}}(x)+\epsilon$
    \item support of the fuzzy subset, $\{x|\mu_{\tilde{X}}(x)>0\}$ is bounded.
\end{enumerate}
\end{definition}

\begin{definition} \cite{dubois1978operations}
The $\nu$ cut of a fuzzy number $\tilde{X}(\nu \in [0,1])$ is a crisp set defined as $\tilde{X}[\nu]=\{x \in \mathbb{R}, \mu_{\tilde{B}}(x) \geq \nu\}$ usually denoted as $\tilde{X}[\nu]=[X_I[\nu],X_{II}[\nu]]$ where $X_I[\nu]=\min \{x \in \mathbb{R}, \mu_{\tilde{B}}(x) \geq \nu\}$ and $X_{II}[\nu]=\max \{x \in \mathbb{R}, \mu_{\tilde{B}}(x) \geq \nu\}$
\end{definition}

\begin{definition}\label{def 1}\cite{buckley2006fuzzy} 
Let $A_i$ be a partition of sample space $X={x_1,x_2,...,x_n}, \tilde{P}(A_i)=\tilde{a_i}$ and $D$ is an event. Then for $\alpha \in [0,1]$ we get
$$\tilde{P}(D)[\nu]=\left\{\sum_{i=1}^na_iP(D|A_i):S\right\}$$ where $S=\{a_i \in \tilde{a_i}[\nu], i \in {1,2,...,n},\sum_{i=1}^na_i=1\}$
\end{definition}
\section{Design of Generalized Fuzzy Multiple Deferred State (GFMDS) Attribute Sampling Plan}
This section aims to review the existing generalized MDS sampling plan for attributes and extend the same in a fuzzy environment. The inspection errors are considered in the next stage and the plan is further extended to inspect the chances of mis-classification. The following are assumptions required for the GMDS as discussed by \cite{aslam2021determination}:
\begin{enumerate}
    \item Continuous inspection is assumed for a series of successive lots.
    \item The quality of lots under consideration is assumed to be uniform.
    \item The destructive nature of testing and the higher cost of inspection makes a small sample size desirable. 
\end{enumerate}
Apart from  the above-stated assumptions, the essential conditions which favor the implementation of the GFMDS plan are:
\begin{enumerate}
    \item The quality level of the lots undergoing examination can be represented as a fuzzy number. In most of the real-time experiments, the determination of exact value is often impossible and fuzzy linguistic quantifiers are used for the representation. 
    \item The reliability of the manufacturing process gives assurance to the consumer. The criteria for acceptance are based on satisfying pre-defined parameters.
    \item The proportion of defectives is observed to follow fuzzy binomial or fuzzy Poisson distribution.
\end{enumerate}

\section{Design of Generalized Fuzzy Multiple Deferred State (GFMDS) Attribute Sampling Plan}
This section aims to review the existing generalized MDS sampling plan for attributes and extend the same in a fuzzy environment. The inspection errors are considered in the next stage and the plan is further extended to inspect the chances of mis-classification. The following are assumptions required for the GMDS as discussed by %\cite{aslam2021determination}:
\begin{enumerate}
    \item Continuous inspection is assumed for a series of successive lots.
    \item The quality of lots under consideration is assumed to be uniform.
    \item The destructive nature of testing and higher cost of inspection makes a small sample size desirable. 
\end{enumerate}
Apart from  the above-stated assumptions, the essential conditions which favor the implementation of the GFMDS plan are:
\begin{enumerate}
    \item The quality level of the lots undergoing examination can be represented as a fuzzy number. In most of the real-time experiments, the determination of exact value is often impossible and fuzzy linguistic quantifiers are used for the representation. 
    \item The reliability of the manufacturing process gives assurance to the consumer. The criteria for acceptance are based on satisfying pre-defined parameters.
    \item The proportion of defectives is observed to follow fuzzy binomial or fuzzy Poisson distribution.
\end{enumerate}
An overview of the GMDS plan is given as:\\
\textbf{Step 1.}
From each lot, a sample of size $n$ is chosen randomly and tested against the required criteria. The percentage of defectives is denoted by $d$.\\
\textbf{Step 2.}
The lot is accepted if $d \leq c_1$ and rejected if $d>c_2$. If $c_1 < d \leq c_2$ the lot under inspection is accepted provided $k$ lots out of $m$ previous lots have been accepted with the condition that $d \leq c_1$. The lot is rejected otherwise. The GMDS plan is differentiated by the parameters $n,k,m,c_1$, and $c_2$. The probability of acceptance of the GMDS sampling plan for attributes is given by
\begin{equation} \label{eq1}
\begin{split}
    P_a(p)=P(d \leq c_1)+P(c_1 < d \leq c_2) \\
    \sum_{j=k}^m \left[{ m \choose {j}} [P(d \leq c_1)]^j[1-P(d \leq c_1)]^{m-j}\right]
\end{split}
\end{equation}
When the fraction of defectives is fuzzy in nature, the existing plan can be extended with fuzzy parameter $\tilde{p}$. 

\subsection{Probability of Acceptance of GFMDS}
The GFMDS plan can be designed so that the proportion of defectives is fuzzy in nature. The assumptions and operating process remains the same except that the random variable $d$ follows a fuzzy probability distribution. \\

\textbf{Theorem 1}\\
The acceptance probability under the GFMDS sampling plan is given by:
\begin{equation*}
\begin{split}
    \tilde{P}_a(\tilde{p})=\tilde{P}(d \leq c_1)+\tilde{P}(c_1 < d \leq c_2)\\
    \sum_{j=k}^m \left[{ m \choose {j}} [\tilde{P}(d \leq c_1)]^j[1-\tilde{P}(d \leq c_1)]^{m-j}\right]
\end{split}
\end{equation*}

\textbf{Proof}:
The number of non-conforming items in a sample is the random variable $d$, which is a crisp value and the fraction of defectives is a triangular fuzzy number, $\tilde{p}=(p_1,p_2,p_3)$. A sample of size $n$ is randomly chosen and inspected. For the unconditional acceptance number $c_1$, if $d \leq c_1$, the lot is accepted. The probability of acceptance will be $\tilde{P}(d \leq c_1)$. For the conditional acceptance number $c_2$, if $c_1 < d \leq c_2$, the current lot is accepted provided the $k$ lots out of $m$ previous lots are accepted in the first step. Therefore, the fuzzy probability of acceptance of $k$ lots out of $m$ lots is determined by 
\begin{equation*}
  \sum_{j=k}^m \left[{ m \choose {j}} [\tilde{P}(d \leq c_1)]^j[1-\tilde{P}(d \leq c_1)]^{m-j}\right]  
\end{equation*}
The fuzzy probability of conditional acceptance can be obtained as:
\begin{equation*}
\begin{split}
    \tilde{P}(c_1 < d \leq c_2)\sum_{j=k}^m \left[{ m \choose {j}}
    [\tilde{P}(d \leq c_1)]^j \right.\\
    \left.[1-\tilde{P}(d \leq c_1)]^{m-j}\right]
    \end{split}
\end{equation*}
Hence, the fuzzy probability of acceptance of the current lot under the proposed GFMDS sampling plan is:
\begin{equation} \label{eq2}
\begin{split}
    \tilde{P}_a(\tilde{p})=\tilde{P}(d \leq c_1)+\tilde{P}(c_1 < d \leq c_2)\\
    \sum_{j=k}^m \left[{ m \choose {j}} [\tilde{P}(d \leq c_1)]^j[1-\tilde{P}(d \leq c_1)]^{m-j}\right]
\end{split}
    \end{equation}
Under  fuzzy Binomial probability distribution, equation (\ref{eq2}) can be written as:
\begin{equation}\label{eq3}
\begin{split}
\tilde{P_a}(\tilde{p}) &= \sum_{d=0}^{c_1} \left[ {n \choose{d}} \tilde{p}^d (1-\tilde{p})^{n-d} \right]+   \\
 &\sum_{d=c_1+1}^{c_2} \left[ {n \choose{d}} \tilde{p}^d (1-\tilde{p})^{n-d}\right]* \\
 &\sum_{j=k}^m {m \choose{j}} \left\{\sum_{d=0}^{c_1} \left[ {n \choose{d}} \tilde{p}^d (1-\tilde{p})^{n-d} \right]\right\}^j *\\
&\left\{1-\sum_{d=0}^{c_1} \left[ {n \choose{d}} \tilde{p}^d (1-\tilde{p})^{n-d} \right]\right\}^{m-j}
\end{split}
\end{equation}
Under fuzzy Poisson probability distribution, equation (\ref{eq2}) can be written as:
\begin{equation}\label{eq4}
\begin{split}
\tilde{P_a}(\tilde{p}) &= \sum_{d=0}^{c_1} \left[ \dfrac{e^{-n\tilde{p}}(n\tilde{p})^d}{d!}\right] + \sum_{d=c_1+1}^{c_2} \left[ \dfrac{e^{-n\tilde{p}}(n\tilde{p})^d}{d!}\right] *  \\
&\sum_{j=k}^m {m \choose j} \left\{\sum_{d=0}^{c_1} \left[ \dfrac{e^{-n\tilde{p}}(n\tilde{p})^d}{d!}\right]\right\}^j \\
&\left\{1-\sum_{d=0}^{c_1} \left[ \dfrac{e^{-n\tilde{p}}(n\tilde{p})^d}{d!}\right]\right\}^{m-j} 
\end{split}
\end{equation}
Here, $\tilde{P_a}(\tilde{p})$ is the fuzzy probability of acceptance for a sample of size $n$ and proportion of defectives $\tilde{p}$ is a fuzzy number. The $\nu$- cut for $\tilde{P_a}(\tilde{p})$ can be defined according to (\cite{buckley2006fuzzy}):\\
Let $\tilde{p}=(p_1,p_2,p_3)$ and $\tilde{q}=(q_1,q_2,q_3)$ be a triangular fuzzy numbers. The $\nu$- cut of $\tilde{p}$ and $\tilde{q}$ is defined as
\begin{equation} \label{eq5}
    \tilde{p}[\nu]=[p_1+(p_2-p_1)\nu, p_3-(p_3-p_2)\nu], \forall 0 \leq \nu \leq 1
\end{equation}
and 
\begin{equation} \label{eq6}
    \tilde{q}[\nu]=[q_1+(q_2-q_1)\nu, q_3-(q_3-q_2)\nu], \forall 0 \leq \nu \leq 1
\end{equation}
Also let
\begin{equation}\label{eq7}
    S=\{p \in \tilde{p}[\nu], q \in \tilde{q}[\nu], p+q=1\}
\end{equation}
The $\nu$- cut for $\tilde{P_a}(\tilde{p})$ can be represented as
\begin{equation} \label{eq8}
    \tilde{P_a}(\tilde{p})[\nu]=[\tilde{P}_{aI}(\tilde{p})[\nu], \tilde{P}_{aII}(\tilde{p})[\nu]] 
\end{equation}
where
\begin{equation} \label{eq9}
    \tilde{P}_{aI}(\tilde{p})[\nu] = \min\{  \tilde{P}_{a}(\tilde{p}):S\} 
\end{equation}
and 
\begin{equation}\label{eq10}
    \tilde{P}_{aII}(\tilde{p})[\nu] =\max \{  \tilde{P}_{a}(\tilde{p}):S\} 
\end{equation}
\subsection{Average Sample Number for GFMDS}
The average sample number of a sampling plan is the average number of units of sample per lot upon which the acceptance or rejection of the lot is made. The average sample number of the proposed GFMDS plan is given as:
\begin{equation*}
    ASN=n
\end{equation*}

\subsection{Average Total Inspection for GFMDS}
Average total inspection is the average number of units per lot which underwent inspection based on the sample size for accepted lots and all units per lot which underwent inspection based on the sample size for rejected lots. ATI of the proposed model can be derived by the following theorem:\\

\textbf{Theorem 2}\\
The ATI for GFMDS plan is given by 
\begin{equation*}
    ATI(\tilde{p})=n+{1-\tilde{P_a}(\tilde{p})}(N-n)
\end{equation*}
where $\tilde{P_a}(\tilde{p})$ is the fuzzy probability of acceptance given by Equation (\ref{eq2}) and $N$ is the lot size.\\
\textbf{Proof}:
The lots submitted are subject to a rectifying inspection process according to the GFMDS plan. \\
\textbf{Case 1.} If there are no non-conformities in the submitted lots, then the whole lot will be accepted. In this case, the number of inspection per lot will be the sample size $n$. \\
\textbf{Case 2.} If the number of non-conformities in the submitted lot are 100\%, then every lot undergoes a rectifying inspection and in that case, the inspected lots will be $N$. \\
\textbf{Case 3.} When the proportion of defectives $\tilde{p}$ is a fuzzy number such that $\tilde{p}[\nu]$ is defined by Equation (\ref{eq5}), the average number of inspected units per lot, ATI will vary between sample size $n$ and lot size $N$. \\
When the fuzzy probability of acceptance for the submitted lots is $\tilde{P_a}(\tilde{p})$ defined by Equation (\ref{eq2}), the ATI per lot will be
\begin{eqnarray*}
\tilde{ATI}(\tilde{p})&=n+\tilde{P_a}(\tilde{p})+{1-\tilde{P_a}(\tilde{p})}N \\
&=N+{1-\tilde{P_a}(\tilde{p})}(N-n) 
\end{eqnarray*}
The $\nu$- cut for $\tilde{ATI}(\tilde{p})$ can be defined according to (\cite{buckley2006fuzzy}):
 \begin{equation*}
 \tilde{ATI}(\tilde{p})[\nu]=[ATI_{I}(\tilde{p})[\nu], ATI_{II}(\tilde{p})[\nu]]
 \end{equation*}
 where
 \begin{equation*}
     ATI_{I}(\tilde{p})=\min \{\tilde{ATI}(\tilde{p}):S\}
 \end{equation*}
 and
 \begin{equation*}
     ATI_{II}(\tilde{p})=\max \{\tilde{ATI}(\tilde{p}):S\}
 \end{equation*}
where $S$ as in Equation (\ref{eq7}).\\

\textbf{Remark 1}
The performance measures of the proposed GFMDS plan, say acceptance probability, ASN and ATI will reduce to the performance measures of FMDS plan defined by \cite{afshari2017designing}
when $k=m$.\\

\textbf{Remark 2}
When $c_1=c_2=c, k=m$ and $m \rightarrow \infty$, the proposed GFMDS plan reduces to the fuzzy single sampling plan for attributes defined by \cite{kahraman2010fuzzy}.

\section{Determination of Optimal Parameters for GFMDS Plan}
The traditional approach for designing an acceptance sampling plan is based on Acceptable Quality Level (AQL) and Limiting Quality Level (LQL), two points on the operating characteristic curve. Such a plan would meet the producer's and consumer's quality requirements. AQL is the worst level of quality for which the consumer would accept the process at an average and LQL is the poorest level of quality that the consumer is willing to accept in an individual lot (\cite{montgomery2009statistical}). In the GMDS plan designed by \cite{aslam2021determination}, the ASN is minimized to obtain the plan parameters (say $n,k,m,c_1$ and $c_2$) for a given pair of $(p_{\alpha},1-\alpha)$ and $(p_{\beta}, \beta)$. Fuzzy ASP are also determined using the above approach. Two points on the FOC band are considered, $(\tilde{p_{\alpha}}, 1-\tilde{\alpha})$ and $(\tilde{p_{\beta}}, \tilde{\beta})$ corresponding to fuzzy AQL (denoted by $\tilde{AQL}$) and fuzzy LQL (denoted by $\tilde{LQL}$) respectively. $\tilde{\alpha}$ is the fuzzy producer's risk and $\tilde{\beta}$ is the fuzzy consumer's risk. The above parameters can be represented as triangular fuzzy numbers as $\tilde{p}$, say,
\begin{eqnarray*}
   \tilde{p_{\alpha}}= &(p_{1\alpha},p_{2\alpha},p_{3\alpha})\\
   \tilde{p_{\beta}}= &(p_{1\beta},p_{2\beta},p_{3\beta})\\
   \tilde{\alpha}=&(\alpha_1, \alpha_2, \alpha_3)\\
   \tilde{\beta}=&(\beta_1, \beta_2, \beta_3)
\end{eqnarray*}
Also, the $\nu$ cut can be obtained as in Equations (\ref{eq5}) and (\ref{eq6}) as:
\begin{equation*}
\begin{split}
\tilde{p_{\alpha}}[\nu] =& [p_{1\alpha}+(p_{2\alpha}-p_{1\alpha})\nu, p_{3\alpha}-(p_{3\alpha}-p_{2\alpha})\nu], \\
&\quad \forall 0 \leq \nu \leq 1 = [p_{\alpha I},p_{\alpha II}]\\
   \tilde{p_{\beta}}[\nu] =& [p_{1\beta}+(p_{2\beta}-p_{1\beta})\nu, p_{3\beta}-(p_{3\beta}-p_{2\beta})\nu], \\
   &\quad \forall 0 \leq \nu \leq 1 = [p_{\beta I},p_{\beta II}]\\
   \tilde{\alpha}[\nu] =& [\alpha_{1}+(\alpha_{2}-\alpha_{1})\nu, \alpha_{3}-(\alpha_{3}-\alpha_{2})\nu], \\
   &\quad \forall 0 \leq \nu \leq 1 = [\alpha_I,\alpha_{II}]\\
   \tilde{\beta}[\nu] =& [\beta_{1}+(\beta_{2}-\beta_{1})\nu, \beta_{3}-(\beta_{3}-\beta_{2})\nu], \\
   &\quad \forall 0 \leq \nu \leq 1 = [\beta_I,\beta_{II}]
   \end{split}
\end{equation*}
Our model assumes that the proportion of defectives is the only parameter of the distribution that is imprecise in nature. All the other parameters are treated according to the classical probability theory. The advances in the design of new sampling schemes intends in diminishing ASN and our proposed plan adopts the same methodology. The solution of the following pair of non-linear optimization problem provides the minimal ASN:
\begin{equation}\label{eq11}
    \tilde{ASN}_I(\tilde{p})=
\begin{cases}
   \min& ASN(\tilde{p})= \tilde{n} \\
   \text{subject to}& \tilde{P_{aI}}(\tilde{p_1}) \geq 1- \tilde{\alpha_I} \\
     &\tilde{P_{aI}}(\tilde{p_2}) \leq \tilde{\beta_I} 
\end{cases}
\end{equation}
\begin{equation} \label{eq12}
 \tilde{ASN}_{II}(\tilde{p})=   
\begin{cases}
   \min& ASN(\tilde{p})= \tilde{n} \\
   \text{subject to}& \tilde{P_{aII}}(\tilde{p_1}) \geq 1- \tilde{\alpha_{II}} \\
   &  \tilde{P_{aII}}(\tilde{p_2}) \leq \tilde{\beta_{II}} 
\end{cases}
\end{equation}
When the value of $\nu=1$, Equations (\ref{eq11}) and (\ref{eq12}) reduces to its crisp counterpart provided by \cite{aslam2021determination}. The ASN becomes a crisp parameter. The results agree with the parameters of the GMDS plan in this case. The pair of non-linear optimization problem reduces to the following problem:
\begin{eqnarray}
   \min & ASN({p})&=n \nonumber\\
   \text{subject to} & P_a({p_1}) &\geq 1-\alpha \\
   & P_a({p_2}) & \leq \beta \nonumber
\end{eqnarray}

Throughout this paper, we assume ASN to be crisp in nature. 
Since our proposed plan consider the plan parameters as crisp values, the ASN will be minimized for a fuzzy proportion of defectives also. Optimal parameters of GFMDS plan for given $\tilde{AQL}$ and $\tilde{LQL}$ for fuzzy binomial probability distribution is given in Table \ref{tab6} and fuzzy Poisson probability model is given in Table \ref{tab7}.

\begin{table*}[htb]
\caption{Optimal parameters of GFMDS plan for fuzzy Binomial distribution when $\tilde{\alpha}[1]=5\%$ and $\tilde{\beta}[1]=10\%$} \label{tab6}
    \begin{tabular}{lcccccc} \hline
        $\tilde{p_1}[1]$& $\tilde{p_2}[1]$& $n$ & $m$ & $k$ & $c_1$ & $c_2$  \\
         \hline
0.001 & 0.01 & 261 & 2 & 1 & 0 & 1 \\
0.0025& 0.05 & 49 & 1 & 1 & 0 & 1 \\
0.005 & 0.10 & 24 & 1 & 1 & 0 & 1 \\
0.0075& 0.08 & 32 & 2 & 1 & 0 & 1 \\
0.010 & 0.04 & 87 & 5 & 1 & 0 & 3 \\
0.015 & 0.05 & 76 & 9 & 1 & 0 & 3 \\
0.02  & 0.08 & 43 & 5 & 1 & 0 & 3 \\
0.05  & 0.15 & 32 & 4 & 1 & 1 & 4 \\
         \hline
    \end{tabular}
    \end{table*}

\begin{table*}[htb]
\caption{Optimal parameters of GFMDS plan for fuzzy Poisson distribution when $\tilde{\alpha}[1]=5\%$ and $\tilde{\beta}[1]=10\%$} \label{tab7}
    \begin{tabular}{lcccccc} \hline
        $\tilde{p_1}[1]$& $\tilde{p_2}[1]$& $n$ & $m$ & $k$ & $c_1$ & $c_2$  \\
         \hline
0.001 & 0.01 & 262 & 2 & 1 & 0 & 1 \\
0.0025& 0.05 & 50 & 1 & 1 & 0 & 1 \\
0.005 & 0.10 & 25 & 1 & 1 & 0 & 1 \\
0.0075& 0.08 & 33 & 2 & 1 & 0 & 1 \\
0.010 & 0.04 & 91 & 6 & 1 & 0 & 3 \\
0.015 & 0.05 & 79 & 10 & 1 & 0 & 3 \\
0.02  & 0.08 & 46 & 6 & 1 & 0 & 3 \\
0.05  & 0.15 & 35 & 6 & 1 & 1 & 4 \\
         \hline
    \end{tabular}
    \end{table*}
    
\textbf{Example 1} \label{eg 1}
Consider a manufacturing system with continuous production. Let the proportion of defectives be a triangular fuzzy number, $\tilde{p}=(0.01,0.02,03)$, which follows a fuzzy Binomial distribution. The GFMDS plan is carried out for screening of the lots with $c_1=0, c_2=3, k=1, m=5$ and $n=87$. For $0 \leq \nu \leq 1$, the $\nu$- cut of $\tilde{p}$ is:
$$\tilde{p}[\nu]=[0.01+0.01 \nu, 0.03-0.01 \nu]$$
$$\tilde{p}[0]=[0.01,0.03]$$
By Equations (\ref{eq5}-\ref{eq10}), the $\nu$- cut for $\tilde{P_a}(\tilde{p})$ can be obtained as:
\begin{equation*}
\begin{split}
\tilde{P_a}(\tilde{p})[\nu]=\left\{\tilde{P}(d \leq 0)+\tilde{P}(0<d\leq 3) \right. \\
\left.\sum_{j=1}^5\left[{5 \choose j}[\tilde{P}(d \leq 0)]^j[1-\tilde{P}(d \leq 0)]^{5-j}\right]:S\right\}
\end{split}
\end{equation*}
For $\nu=0$, we get
$$\tilde{P_a}(\tilde{p})[0]=[0.28,0.95]$$
This means for all 100 lots underwent an inspection process, 28 to 95 lots will be accepted with at least degree of membership zero. Consider the case when $p$ is crisp. The acceptance probability in this case can be evaluated by taking $\nu=1$, which gives $p=0.02$ and $P_a=0.62$. When the rate of uncertainty is expressed in terms of linguistic fuzzy parameters, say the chance that the value of $p=0.02$ is 30\%, we can express this in terms of a fuzzy number by considering $\nu=0.3$. In this case, we obtain $\tilde{p}[\nu]=[0.013,0.027]$ and $\tilde{P_a}(\tilde{p})[\nu]=[0.36,0.87]$. That is, for all 100 lots underwent an inspection process, 36 to 87 lots will be accepted as per the inspection policy.

\textbf{Example 2} \label{eg 2}
The proposed GFMDS plan can be applied to a manufacturing system where the proportion of non conformity's follow a fuzzy Poisson distribution. Let the fraction of defectives be a triangular fuzzy number, say $\tilde{p}=(0.02,0.03,0.04)$ and the sampling plan parameters are: $n=86,c_1=1,c_2=4,k=1,m=5$. For $0 \leq \nu \leq 1$, the $\nu$- cut of $\tilde{p}$ is given by:
$$\tilde{p}[\nu]=[0.02+0.01 \nu, 0.04-0.01 \nu]$$
For $\nu=0$, $\tilde{p}[\nu]=[0.02,0.04]$. Using Equations (\ref{eq7}-\ref{eq10}), for $\nu=0$, we get
$$\tilde{P_a}(\tilde{p})[0]=[0.57,0.95]$$
That is, for every 100 lots in the inspection process, it is expected that 57 to 95 lots are accepted with the degree of membership zero. When $\nu=1$, the plan reduces to the classical MDS plan with fraction of defectives $p=0.03$ (Since $\tilde{p}[1]=[0.02+0.01*1,0.04-0.01*1]=0.03$). In this case, the probability of acceptance will be $P_a(p)=0.79$. 

\begin{figure}
\centering
{
{\includegraphics[width= 7 cm, height = 7 cm]{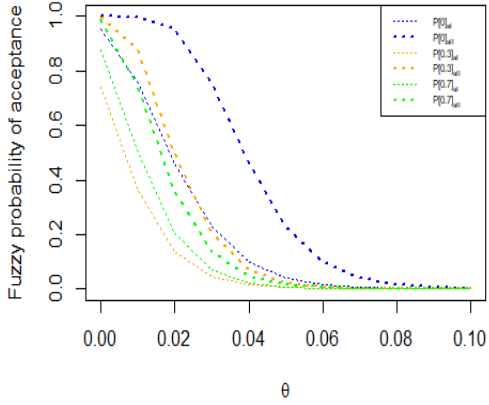}}}%\hspace{10pt}
\caption{FOC bands for GFMDS plan for different values of $\nu$} \label{fig1}
\end{figure}

\textbf{Example 3} \label{eg 3}
In this example, the FOC band is obtained for the parameters in Example (\ref{eg 2}). The assumptions remains unchanged. Since $\tilde{p}$ is a fuzzy random variable, the probability of acceptance $\tilde{P_a}$ is also fuzzy. The fuzzy OC band is obtained in this case (Figure \ref{fig1}) unlike the OC curve when $p$ and $P_a$ are crisp values. In order to construct the FOC band, we consider a facilitator parameter $\theta$ to identify the variations of $\tilde{p}$ (see \cite{afshari2017designing}).
Now $\tilde{p}=(a_1,a_2,a_3)$ can be written as $\tilde{p_\theta}$ as follows:
$$\tilde{p_\theta}=(\theta,b_2+\theta,b_3+\theta)$$
$$\tilde{p_\theta}[\nu]=[\theta+b_2\nu,b_3+\theta-(b_3-b_2)\nu], \quad \forall 0 \leq \nu \leq 1$$
where $b_j=a_j-a_1, \quad(j=2,3)$ and $\theta \in [0,1-b_3]$. The probability of acceptance for $\tilde{p_{\theta}}$ can be obtained by replacing $p$ in Equations (\ref{eq2}) and (\ref{eq4}) by $p_{\theta}$. For $\tilde{p}=(0.02,0.03,0.04)$, we have $\tilde{p_\theta}=(\theta,0.01+\theta,0.02+\theta)$ and 
\begin{equation*}
\begin{split}
\tilde{p_\theta}[0]&=(\theta+0.01*0,0.02+\theta-0.01*0], \\
&=[\theta,0.02+\theta]\quad \forall 0 \leq \theta \leq 0.98
\end{split}
\end{equation*}
The fuzzy probability of acceptance for different values of $\theta$ when $\nu=0$ is given in Table (\ref{tab1}).

\begin{table*}[htb]
\caption{The fuzzy probability of lot acceptance for GFMDS plan without inspection errors corresponding to different values of $\nu$ and $\theta$} \label{tab1}
    {\begin{tabular}{lccccccc} \hline
$\theta$ & $\tilde{p_{\theta}}$ & $\tilde{p_\theta}[0]$ & $\tilde{P_a}(\tilde{p_\theta})[0]$ & $\tilde{p_\theta}[0.3]$ & $\tilde{P_a}(\tilde{p_\theta})[0.3]$ & $\tilde{p_\theta}[0.7]$ & $\tilde{P_a}(\tilde{p_\theta})[0.7]$\\  \hline
0&(0,0.01,0.02)&[0,0.02]&[0.952,1.00]&[0.003,0.017]&[0.1411,0.7391]&[0.007,0.013]&[0.8781,0.9876]\\
0.01&(0.01,0.02,0.03)&[0.01,0.03]&[0.755,0.998]&[0.013,0.027]&[0.3608,0.8781]&[0.017,0.023]&[0.5011,0.7391]\\
0.02&(0.02,0.03,0.04)&[0.02,0.04]&[0.461,0.952]&[0.023,0.037]&[0.1352,0.5011]&[0.027,0.033]&[0.2049,0.3608]\\
0.03&(0.03,0.04,0.05)&[0.03,0.05]&[0.227,0.755]&[0.033,0.047]&[0.0444,0.2049]&[0.037,0.043]&[0.0698,0.1352]\\
0.04&(0.04,0.05,0.06)&[0.04,0.06]&[0.098,0.461]&[0.043,0.057]&[0.0138,0.0698]&[0.047,0.053]&[0.0221,0.0444]\\
0.05&(0.05,0.06,0.07)&[0.05,0.07]&[0.039,0.227]&[0.053,0.067]&[0.0042,0.0221]&[0.057,0.063]&[0.0068,0.0138]\\
0.06&(0.06,0.07,0.08)&[0.06,0.08]&[0.015,0.098]&[0.063,0.077]&[0.0013,0.0068]&[0.067,0.073]&[0.0022,0.0042]\\
0.07&(0.07,0.08,0.09)&[0.07,0.09]&[0.006,0.039]&[0.073,0.087]&[0.0005,0.0022]&[0.077,0.083]&[0.0006,0.0013]\\
0.08&(0.08,0.09,0.10)&[0.08,0.10]&[0.002,0.015]&[0.083,0.097]&[0,0.0006]&[0.087,0.093]&[0.0002,0.0005]\\
0.09&(0.09,0.10,0.11)&[0.09,0.11]&[0.001,0.006]&[0.093,0.107]&[0,0.0002]&[0.097,0.103]&[0,0.0002]\\
0.10&(0.10,0.11,0.12)&[0.10,0.12]&[0.001,0.002]&[0.103,0.117]&[0,0]&[0.107,0.113]&[0,0]\\ 
 \hline
\end{tabular}}
    \end{table*}
    
The next section validates the efficiency of the GFMDS plan by showing that the ASN is minimum for the proposed plan on comparison with the existing fuzzy acceptance sampling plans. 

\section{Probability of acceptance of GFMDS Plan in terms of generalised fuzzy number}
Throughout this paper, the fraction of defectives is assumed to be a triangular fuzzy number. However, this can be generalised to a fuzzy number with a different membership function. Let $f_{z_i}(\tilde{p})$ be the probability density function for fuzzy proportion of defectives $\tilde{p}$ for parameters $z_i$ where $i=1,2,...,n$. If $p$ and $\mathbb{P}$ denotes the crisp proportion of defectives and crisp probability measure respectively, we can express $\tilde{p}$ as $$\tilde{p}=\{(p,\mu_{\tilde{p}}(p)|p \in \mathbb{P}\}$$ where $\mu_{\tilde{p}}(p)$ denotes the membership function of $\tilde{p}$. The $\nu$- cut or $\nu$ level sets of $z_i$ is 
\begin{equation*}
\begin{aligned}
\displaystyle \tilde{z_i}[\nu]&=\left[\underset{z_i}{\min} \{z_i \in S(\tilde{z_i})|\mu_{\tilde{z_i}}(z_i) \geq \nu \}, \right. \\
 &\left.\underset{z_i}{\max} \{z_i \in S(\tilde{z_i})|\mu_{\tilde{z_i}}(z_i) \geq \nu \} \right]\\
&=[\tilde{z_{iI}}[\nu],\tilde{z_{iII}}[\nu]]
\end{aligned}
\end{equation*}
where $S(\tilde{z_i})=\{z_i \in \tilde{z_i}[\nu], \sum_{i=1}^n z_i=1\}$. \\
The upper and lower bound of $\nu$- cut of $\tilde{P_a}(\tilde{p})$ can be found by solving the following pair of optimization problem:
$$\tilde{P_{aI}}(\tilde{p})[\nu] =
\begin{cases}
\min P_a(p) \\
\text{s.t.} \tilde{z_{iI}}[\nu] \leq \tilde{z_i} \leq \tilde{z_{iII}}[\nu]
\end{cases}$$
and
$$\tilde{P_{aII}}(\tilde{p})[\nu] =
\begin{cases}
\max P_a(p) \\
\text{s.t.} \tilde{z_{iI}}[\nu] \leq \tilde{z_i} \leq \tilde{z_{iII}}[\nu]
\end{cases}$$
where $P_a(p)$ denotes the conventional probability of acceptance when $p$ is crisp. Thus $\nu$- cut of $\tilde{P_{a}}(\tilde{p})$ is $$\tilde{P_{a}}(\tilde{p})[\nu]=[\tilde{P_{aI}}(\tilde{p})[\nu],\tilde{P_{aII}}(\tilde{p})[\nu]]$$
According to \cite{zimmermann2011fuzzy}, the $\nu$- cuts form a nested interval for different values of $\nu$. For $0<\nu_1 < \nu_2 \leq 1$, we have $\tilde{P_{aI}}(\tilde{p})[\nu_2]\geq \tilde{P_{aI}}(\tilde{p})[\nu_1]$ and $\tilde{P_{aII}}(\tilde{p})[\nu_1]\geq \tilde{P_{aII}}(\tilde{p})[\nu]$

\textbf{Example 4}\label{eg6}
Consider the proportion of defectives having a pentagonal membership function and following a fuzzy Binomial distribution. The $\nu$- cut for a pentagonal fuzzy number $\tilde{x}=(x_1,x_2,x_3,x_4,x_5)$ is $[2 \nu (x_2-x_1)+x_1,-2 \nu (x_5-x_4)+x_5]$
for $\nu \in [0,0.5]$ and $[2\nu (x_3-x_2)+2x_2-x_3, 2\nu(x_4-x_3)-x_4+2x_3]$
for $\nu \in [0.5,1]$.
Let the fuzzy proportion of defectives be $\tilde{p}=(0.02,0.03,0.05,0.07,0.08)$ and plan parameters be $n=87,c_1=0,c_2=3,k=1,m=5$. Then, $$
\tilde{p}[\nu]=
\begin{cases}
\begin{aligned}
[0.02+0.02 \nu, 0.08-0.02 \nu], & \quad \nu \in [0,0.5]\\
[0.01+0.04 \nu, 0.17-0.04 \nu], & \quad \nu \in [0.5,1]
\end{aligned}
\end{cases}
$$
The probability of acceptance when $p$ is a crisp variable is given in Equation (\ref{eq1}) \\
\textit{Case 1}: $\nu \in [0,0.5]$\\
To find $\nu$- cut for $\tilde{P_a}(\tilde{p})$, consider the following pair of optimization problems:
$$\tilde{P_{aI}}(\tilde{p})=
\begin{cases}
\min \quad P_a(p) \\
\text{s.t.}\quad 0.02+0.02 \nu \leq p \leq 0.08-0.02 \nu
\end{cases}$$
and 
$$\tilde{P_{aII}}(\tilde{p})=
\begin{cases}
\max \quad P_a(p) \\
\text{s.t.}\quad 0.02+0.02 \nu \leq p \leq 0.08-0.02 \nu
\end{cases}$$
Solving the above problems, the $\nu$- cut of fuzzy probability of acceptance for say $\nu=0.3$ is obtained as:
$\tilde{P_a}(\tilde{p})[0.3]=[0.002,0.393]$\\
\textit{Case 2}: $\nu \in [0.5,1]$\\
To find $\nu$- cut for $\tilde{P_a}(\tilde{p})$, consider the following pair of optimization problems:
$$\tilde{P_{aI}}(\tilde{p})=
\begin{cases}
\min \quad P_a(p) \\
\text{s.t.}\quad 0.01+0.04 \nu \leq p \leq 0.17-0.04 \nu
\end{cases}$$
and 
$$\tilde{P_{aII}}(\tilde{p})=
\begin{cases}
\max \quad P_a(p) \\
\text{s.t.} \quad 0.01+0.04 \nu \leq p \leq 0.17-0.04 \nu
\end{cases}$$
Solving the above problems, the $\nu$- cut of fuzzy probability of acceptance for say, $\nu=0.8$ is obtained as:
$\tilde{P_a}(\tilde{p})[0.8]=[0,0.27]$

\section{Efficiency of GFMDS Plan}
The limitation of classical MDS plans is that the lots are being placed in a state subject to certain conditions for an indefinite time. The traditional single sampling plan has the limitation that a lot is rejected as soon as a defective item is identified \cite{aslam2021determination} showed that a GMDS plan overcomes the above-said shortcomings and safeguards the interests of the producer and consumer. They also showed that the GMDS plan is designed with minimum ASN and ATI compared to the existing sampling schemes. In this section, the authors would like to investigate the efficiency of the proposed GFMDS plan over the existing fuzzy acceptance sampling plans for attributes. We compare the GFMDS plan with fuzzy SSP and fuzzy DSP proposed by \cite{kahraman2010fuzzy} and fuzzy MDS plan designed by \cite{afshari2017designing}. The ASN for each of the sampling schemes is obtained for a combination of $\tilde{AQL}$ and $\tilde{LQL}$. The results obtained for fuzzy Binomial distribution are shown in Table (\ref{tab 4}) and for fuzzy Poisson distribution is shown in Table (\ref{tab 5}). From the tables, it is clear that the ASN has been considerably reduced when the GFMDS plan is employed which proves the economic advantage of GFMDS plan over the existing fuzzy acceptance sampling plans.

\begin{table*}[htb]
\caption{ASN of FSSP, FDSP, FMDS and GFMDS for fuzzy Binomial distribution} \label{tab 4}
   {\begin{tabular}{lccccc} \hline
$\tilde{p_1}[1]$& $\tilde{p_2}[1]$& FSSP & FDSP & FMDS & GFMDS \\ \hline
0.001 & 0.010 & 531 & 363.55 & 388 & 261 \\
0.001 & 0.015 & 258 & 188.11 & 165 & 165 \\
0.0025& 0.005 & 4948& 4942.89&3197 & 1822 \\
0.005 & 0.010 & 2473& 2471.44&1597 & 910 \\
0.010 & 0.030 & 390 & 364.49 & 275 & 170 \\
0.010 & 0.050 & 132 & 103.16 & 87  & 62 \\
 \hline
\end{tabular}}
    \end{table*}

\begin{table*}[htb]
\caption{ASN of FSSP, FDSP, FMDS and GFMDS for fuzzy Poisson distribution} \label{tab 5}
   {\begin{tabular}{lccccc} \hline
$\tilde{p_1}[1]$& $\tilde{p_2}[1]$& FSSP & FDSP & FMDS & GFMDS \\ \hline
0.001 & 0.010 & 533 & 364.97 & 390 & 262 \\
0.001 & 0.015 & 260 & 190.59 & 167 & 167 \\
0.0025& 0.005 & 4952& 4946.86&3200 & 1824 \\
0.005 & 0.010 & 2476& 2473.43&1600 & 912 \\
0.010 & 0.030 & 393 & 366.33 & 277 & 172 \\
0.010 & 0.050 & 134 & 127.76 & 108  & 63 \\
 \hline
\end{tabular}}
    \end{table*}

\begin{table*}[htb]
\caption{Comparison of fuzzy probability of lot acceptance for GFMDS plan with and without inspection errors for $\nu=0$} \label{tab 3}
   {\begin{tabular}{lcccc} \hline
$\theta$ &$\tilde{p_{\theta}}[0]$ & $\tilde{P_a}(\tilde{p_{\theta}})[0]$ & $\tilde{p_{\delta\theta}}[0]$ & $\tilde{P_a}(\tilde{p_{\delta\theta}})[0]$\\  \hline
0&[0,0.2]&[0.952,1.00]&[0,0.0182]&[0.6917,1.00]\\
0.01&[0.01,0.03]&[0.755,0.998]&[0.01,0.0282]&[0.3243,0.95]\\
0.02&[0.02,0.04]&[0.461,0.952]&[0.02,0.0382]&[0.1190,0.6193]\\
0.03&[0.04,0.06]&[0.227.0.755]&[0.03,0.0482]&[0.0386,0.2745]\\
0.04&[0.05,0.07]&[0.098,0.461]&[0.04,0.0582]&[0.0119,0.0978]\\
0.05&[0.06,0.08]&[0.039,0.227]&[0.05,0.0682]&[0.0037,0.0313]\\
0.06&[0.07,0.09]&[0.015,0.098]&[0.06,0.0782]&[0.0012,0.0096]\\
 \hline
\end{tabular}}
    \end{table*}
    
\section{GFMDS plan with Inspection Errors}
One of the blind assumptions in acceptance sampling plans is the absence of any measurement error. The traditional sampling plans were carried out in this belief which in turn affects the system reliability as a whole. The possible errors in an inspection process is of two types: a perfect item is classified as an imperfect item and vice versa. These errors are named as Type I error and Type II error respectively (hereafter denoted by $\delta_1$ and $\delta_2$). The GFMDS plan with inspection error is designed by considering the possible events of inspection process:\\
$A_1$: The item is imperfect\\
$A_2$: The item is perfect\\
$B$: The item is classified as imperfect post-inspection\\
$B|A_2$: A perfect item is classified as imperfect\\
$B^c|A_1$: An imperfect item is classified as perfect\\
The actual and experimental values of fuzzy proportion of defectives are denoted by $\tilde{p}$ and $\tilde{p_{\delta}}$. Then, $\tilde{p}=\tilde{P}(A_1)$ and $\tilde{p_e}=\tilde{P}(B)$. Also, $\delta_1=P(B|A_2)$ and $\delta_2=P(B^c|A_1)$. By Definition \ref{def 1}, for $\nu \in [0,1],$
\begin{equation}\label{eq14}
    \tilde{p_{\delta}}=\tilde{p}(1-\delta_2)+\tilde{q}\delta_1
\end{equation}
The $\nu$- cut is given by
\begin{equation} \label{eq15}
    \tilde{p_{\delta}}[\nu]=\{\tilde{p}(1-\delta_2)+\tilde{q}\delta_1:S\}
\end{equation}
where $S$ is defined by:
\begin{equation} \label{eq16}
 S=\{p \in \tilde{p_{\delta}}[\nu],q \in \tilde{q_{\delta}}[\nu],p_{\delta}+q_{\delta}=1\}   
\end{equation}
The fuzzy probability of acceptance when inspection errors are considered can be obtained by replacing $\tilde{p}$ by $\tilde{p_{\delta}}$ in Equations (\ref{eq3}) and (\ref{eq4}) for Binomial and Poisson models respectively for $S$ given in Equation (\ref{eq16}). The $\nu$- cut is obtained similarly which is represented as
\begin{equation} \label{eq17}
    \tilde{P_a}(\tilde{p_{\delta}})[\nu] =[\tilde{P}_{eI}[\nu], \tilde{P}_{eII}[\nu]]
\end{equation}
throughout this article.

\textbf{Example 5} \label{eg 4}
Suppose that the inspection process carried out in the production system in Example (\ref{eg 1}) is found to be imperfect. The inspection errors are identified as $\delta_1=0.01$ and $\delta_2=0.08$ from the previous lots which were subjected to the same inspection process. If all the process parameters remains unchanged, then the experimental fuzzy proportion of defectives were obtained as:
$$\tilde{p_{\delta}}=(0.0191,0.0282,0373)$$
$$\tilde{p}[\nu]=[0.01+0.01 \nu,0.03-0.01 \nu]$$
$$\tilde{p_{\delta}}=[0.0191+0.0091 \nu,0.0373-0.0091 \nu]$$
When $\nu=0$,
$$\tilde{p}[0]=[0.01,0.03] \quad \text{and} \quad \tilde{p_{\delta}}[0]=[0.0191,0.0373]$$
Using Equations (\ref{eq15} - \ref{eq17}), the 0- cut of fuzzy probability of lot acceptance with inspection errors is:
$$\tilde{P_a}(\tilde{p_{\delta}})[0]=[0.13,0.66]$$ It is expected that for every 100 lots, 13 to 66 lots are accepted with a least degree of membership zero. Comparing with GFMDS plan as in Example \ref{eg 1}, it is seen that the number of accepted lot reduces when inspection errors are considered. The fuzzy probability of acceptance for different values of inspection errors are shown in Table \ref{tab1}.

\begin{figure}
\centering
{{\includegraphics[width= 7 cm, height = 7 cm]{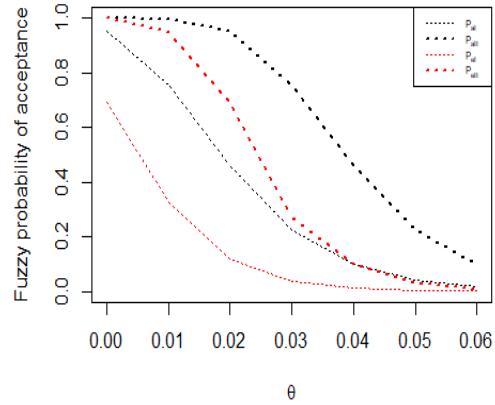}}}%\hspace{10pt}
\caption{FOC bands for GFMDS plan with and without inspection errors} \label{fig2}
\end{figure}

\textbf{Example 6} \label{eg 5}
The effect of inspection errors on GFMDS plan can be studied by constructing and comparing FOC bands for GFMDS plan with and without inspection errors. As in Example \ref{eg 3}, a facilitator parameter $\theta$ can be used to plot the FOC bands. Consider the assumptions in Example \ref{eg 4}. $\tilde{p_{\delta}}=(c_1,c_2,c_3)$ can be represented using the facilitator parameter $\theta$ as:
$$\tilde{p_{\delta\theta}}=(\theta, d_2+\theta, d_3+\theta)$$
The $\nu$- cut of $\tilde{p_{\delta\theta}}$ is:
$$\tilde{p_{\delta\theta}}[\nu]=[\theta+d_2+\theta,d_3+\theta-(d_3-d_2)\alpha], \quad \forall 0 \leq \nu \leq 1$$
where $d_i=c_i-c_1 \quad (i=2,3)$ and $\theta \in [0,1-d_3]$. For finding the probability of acceptance, the fuzzy proportion of defectives in equations %(\ref{eq 7}-\ref{eq 9})
can be replaced with $\tilde{p_{\delta\theta}}$. For $\tilde{p_{\delta}}=(0.0191,0.0282,0.0373)$, $\tilde{p_{\delta\theta}}=[\theta, 0.0091+\theta,0.0182+\theta]$. The $\nu$ cut of $\tilde{p_{\delta\theta}}$ is:
$$\tilde{p_{\delta\theta}}[\nu]=[\theta+ 0.0091\nu,0.0182+\theta-0.0091\nu]$$ For $\nu=0,$ 
$$\tilde{p_{\delta\theta}}[0]=[\theta, 0.0182+\theta] \quad \forall 0 \leq \theta \leq 0.9818 $$ 
The values of $\tilde{p_{\delta\theta}}[0]$ and $\tilde{P_a}(\tilde{p_{\delta}})[0]$ are shown in Table (\ref{tab 3}) for different values of $\theta$. Also, the FOC band for GFMDS plan with and without inspection errors is constructed (see, Figure \ref{fig2}). The results shows that the fuzzy probability of acceptance with and without considering the measurement errors is falling when the 0- cut of the experimental and actual values of fuzzy proportion of defectives are rising.

\section{Conclusion}
This study extended the generalized MDS sampling plan to the fuzzy environment. The uncertainty in determining the exact value of the fraction of defectives is dealt by the fuzzy set theory. The design of GFMDS plan is reduced to a classical GMDS plan when $p$ is crisp and thus the plan is well-defined. The numerical examples illustrates the construction, advantage and effect of errors on the proposed plan. The efficiency of GFMDS is established by comparing the average sample number with that of single, double, and multiple deferred plans in the fuzzy environment. The plan is rephrased including inspection errors to deal with the imperfect measurement process. The drop in the probability of acceptance is identified when sampling errors are identified. The obtained results show this plan overcame the limitations of the existing fuzzy acceptance sampling plans.

\subsection*{Acknowledgement}
The authors would like to thank DST, Govt. of India for extending the laboratory support under the project  (SR/FST/MS-1/2019/40) of Department of Mathematics, NIT Calicut. The first author also like to thank CSIR, Govt. of India for extending financial support (09/874(0039)/2019-EMR-I).

%%%%%%%%%%% The bibliography starts:


\begin{thebibliography}{9}
\bibitem{afshari2017construction}
R. Afshari and B.S. Gildeh, Construction of fuzzy multiple deferred state sampling plan, \textit{2017 Joint 17th World Congress of International Fuzzy Systems Association and 9th International Conference on Soft Computing and Intelligent Systems (IFSA-SCIS) IEEE} (2017a), 1--7. 
\bibitem{afshari2017modified}
R. Afshari and B.S. Gildeh, Modified sequential sampling plan using fuzzy sprt, \textit{2017 5th Iranian Joint Congress on Fuzzy and Intelligent Systems (CFIS), IEEE} (2017b), 116--121. 
\bibitem{afshari2017designing}
R. Afshari and B.S. Gildeh, Designing a multiple deferred state attribute sampling plan in a fuzzy environment, \textit{American Journal of Mathematical and Management Sciences} \textbf{36}(4) (2017), 328--345. 
\bibitem{afshari2018effects}
R. Afshari and B.S. Gildeh, The effects of misclassification errors on multiple deferred state attribute sampling plan, \textit{Journal of Industrial and Systems Engineering} \textbf{11}(2) (2018a), 31--46. 
\bibitem{afshari2018fuzzy}
R. Afshari and B.S. Gildeh, Fuzzy multiple deferred state variable sampling plan, \textit{Journal of Intelligent \& Fuzzy Systems} \textbf{34}(4) (2018b), 2737--2752. 
\bibitem{afshari2017fuzzy}
R. Afshari, B.S. Gildeh and M. Sarmad, Fuzzy multiple deferred state attribute sampling plan in the presence of inspection errors, \textit{Journal of Intelligent \& Fuzzy Systems} \textbf{33}(1) (2017), 503--514. 
\bibitem{anderson2001acceptance}
M. T. Anderson, B.S. Greenberg and S. L. Stokes, Acceptance sampling with rectification when inspection errors are present, \textit{Journal of Quality Technology} \textbf{33}(4) (2001), 493--505.
\bibitem{aslam2021determination}
M. Aslam, S. Balamurali, and C. H. Jun, Determination and economic design of a generalized multiple dependent state sampling plan, \textit{Communications in Statistics-Simulation and Computation} \textbf{50}(11) (2021a), 3465--3482.
\bibitem{aslam2021new}
M. Aslam, S. Balamurali, and C. H. Jun, A new multiple dependent state sampling plan based on the process capability index, \textit{Communications in Statistics-Simulation and Computation} \textbf{50}(6) (2021b), 3465--3482.
\bibitem{aslam2015new}
M. Aslam, A. Nazir, and C. H. Jun, A new attribute control chart using multiple dependent state sampling, \textit{Transactions of the Institute of Measurement and Control} \textbf{37}(4) (2015), 3465--3482.
\bibitem{aslam2014multiple}
M. Aslam, C.-H. Yen, C.-H. Chang and C. H. Jun, Multiple dependent state variable sampling plans with process loss consideration, \textit{The International Journal of Advanced Manufacturing Technology} \textbf{71}(5) (2014), 1337--1343
\bibitem{balamurali2007multiple}
S. Balamurali, and C. H. Jun, Multiple dependent state sampling plans for lot acceptance based on measurement data, \textit{European Journal of Operational Research} \textbf{180}(3) (2007), 1221--1230
\bibitem{baloui2011inspection}
B. E. Jamkhaneh, B. S. Gildeh, and G. Yari, Inspection error and its effects on single sampling plans with fuzzy parameters, \textit{Structural and multidisciplinary Optimization} \textbf{43}(4) (2011), 555--560
\bibitem{bhattacharya2020generalized}
R. Bhattacharya,and M. Aslam, Generalized multiple dependent state sampling plans in presence of measurement data, \textit{IEEE Access} \textbf{8} (2020), 162775--162784
\bibitem{buckley2006fuzzy}
J. J. Buckley, \textit{Fuzzy Probability and Statistics}, Vol. 196, Berlin: Springer, 2006. 
\bibitem{chakraborty1992class}
T. Chakraborty, A class of single sampling plans based on fuzzy optimisation, \textit{Quality control and applied statistics} \textbf{37}(7) (1992), 359--362
\bibitem{collins1973effects}
R. D. Collins Jr, K. E. Case and G. K. Bennett, The effects of inspection error on single sampling inspection plans, \textit{International Journal of Production Research} \textbf{11}(3) (1973), 289--298
\bibitem{dodge1977chain}
H. Dodge, Chain sampling inspection plans, \textit{Journal of Quality Technology} \textbf{9}(3) (1977), 139--142
\bibitem{dodge1929method}
H. F. Dodge, and H. G. Romig, A method of sampling inspection, \textit{The Bell System Technical Journal} \textbf{8}(4) (1929), 613--631
\bibitem{dodge1941single}
H. F. Dodge, and H. G. Romig, Single sampling and double sampling inspection tables, \textit{The Bell System Technical Journal} \textbf{20}(1) (1941), 1--61
\bibitem{dubois1978operations}
D. Dubois, and H. Prade, Operations on fuzzy numbers, \textit{International Journal of systems science} \textbf{9}(6) (1978), 613--626
\bibitem{fard1993analysis}
N. S. Fard, and J. J. Kim, Analysis of two stage sampling plan with imperfect inspection, \textit{Computers \& Operations Research} \textbf{25}(1--4) (1993), 453--456
\bibitem{ferrell2002design}
W. G. Ferrell Jr, and A. Chhoker, Design of economically optimal acceptance sampling plans with inspection error, \textit{Computers \& Operations Research} \textbf{29}(10) (2002), 1283--1300
\bibitem{govindaraju2007inspection}
K. Govindaraju, Inspection error adjustment in the design of single sampling attributes plan, \textit{Quality Engineering} \textbf{19}(3) (2007), 227--233.
\bibitem{govindaraju1993selection}
K. Govindaraju, and K. Subramani, Selection of multiple deferred (dependent) state sampling plans for given acceptable quality level and limiting quality level, \textit{Journal of Applied Statistics} \textbf{20}(3) (1993), 423--428.
\bibitem{grzegorzewski2001acceptance}
P. Grzegorzewski, \textit{Acceptance sampling plans by attributes with fuzzy risks and quality levels}, in: Frontiers in Statistical Quality Control 6, Springer, \textbf{20}(3) 2001, 36--46.
\bibitem{isik2021design}
G. Isik and I. Kaya, Design and analysis of acceptance sampling plans based on intuitionistic fuzzy linguistic terms, \textit{Iranian Journal of Fuzzy Systems} \textbf{18}(6) (2021), 101--118.
\bibitem{kahraman2016fuzzy}
C. Kahraman, E. T. Bekar, and O. Senvar, A fuzzy design of single and double acceptance sampling plans, \textit{Intelligent decision making in quality management, Springer} (2016), 179--211.
\bibitem{kahraman2010fuzzy}
C. Kahraman, and I. Kaya, \textit{Fuzzy acceptance sampling plans}, in Production engineering and management under fuzziness, Springer, 2010, 457--481.
\bibitem{kuralmani1992selection}
V. Kuralmani, and K. Govindaraju, Selection of multiple deferred (dependent) state sampling plans, \textit{Communications in statistics-theory and methods}, \textbf{21}(5) (1992), 1339--1366
\bibitem{montgomery2009statistical}
D. C. Montgomery, \textit{Statistical quality control}, Volume 7, Wiley New York, 2009.
\bibitem{sadeghpour2011acceptance}
B. S. Gildeh, B. E. Jamkhaneh, and G. Yari, Acceptance single sampling plans with fuzzy parameters, \textit{Iranian Journal of Fuzzy Systems} \textbf{8}(2) (2011), 47--55
\bibitem{soundararajan1990construction}
V. Soundararajan, and R. Vijayaraghavan, Construction and selection of multiple dependent (deferred) state sampling plan, \textit{Journal of Applied Statistics} \textbf{17}(3) (1990), 397--409
\bibitem{turanouglu2012fuzzy}
E. Turanoglu, I. Kaya, and C. Kahraman, Fuzzy acceptance sampling and characteristic curves, \textit{International Journal of Computational Intelligence Systems} \textbf{5}(1) (2012), 13--29
\bibitem{wortham1976multiple}
A. Wortham and R. Baker, Multiple deferred state sampling inspection, \textit{The International Journal of Production Research} \textbf{14}(6) (1976), 719--731
\bibitem{wortham1970dependent}
A. Wortham and J. Mogg, Dependent stage sampling inspection, \textit{The International Journal of Production Research} \textbf{8}(4) (1970), 385--395
\bibitem{wu2017developing}
C. W. Wu and Z. H. Wang, Developing a variables multiple dependent state sampling plan with simultaneous consideration of process yield and quality loss, \textit{The International Journal of Production Research} \textbf{55}(8) (2017), 2351--2364
\bibitem{yan2016designing}
A. Yan, S. Liu, and X. Dong, Designing a multiple dependent state sampling plan based on the coefficient of variation, \textit{SpringerPlus} \textbf{5}(1) (2016), 1--13
\bibitem{zimmermann2011fuzzy}
H. J. Zimmermann, \textit{Fuzzy set theory—and its applications}, Springer Science \& Business Media, 2001.
\end{thebibliography}
\end{document}